\documentclass[aps,prd,reprint,preprintnumbers,floats,epsfig,nofootinbib,amssymb,
nofootinbib]{revtex4}

\usepackage{slashed}
\usepackage{graphicx,color}
\usepackage{epsfig}
\usepackage{subfigure}
\usepackage{epsfig}
\usepackage{epstopdf}
\usepackage{dcolumn}
\usepackage{bm}
\usepackage{color}
\usepackage{ulem}
\usepackage{enumitem}
\usepackage[colorlinks,
            linkcolor=black,
            anchorcolor=black,
            citecolor=black
            ]{hyperref}

\begin{document}

\baselineskip=15pt


\title{ $SU(3)$ Flavor Symmetry for Weak Hadronic Decays of ${\bf B}_{bc}$ Baryons}

\author{Junxing Pan$^{1}$\footnote{panjunxing2007@163.com}}
\author{Yu-Kuo Hsiao$^{1}$\footnote{yukuohsiao@gmail.com}}
\author{Jin Sun${}^{2}$\footnote{019072910096@sjtu.edu.cn}}
\author{Xiao-Gang He${}^{3,4}$\footnote{hexg@phys.ntu.edu.tw}}

\affiliation{${}^{1}$School of Physics and Information Engineering, Shanxi Normal University, Linfen 041004, China}
\affiliation{${}^{2}$Tsung-Dao Lee Institute, SKLPPC, MOE KLPPC, and School of Physics and Astronomy, Shanghai Jiao Tong University, Shanghai 200240, China}
\affiliation{${}^{3}$Department of Physics, National Taiwan University, Taipei 10617, Taiwan}
\affiliation{${}^{4}$Physics Division, National Center for Theoretical Sciences, Hsinchu 30013, Taiwan}

\begin{abstract}
Baryons with a heavy c-quark or a heavy b-quark and also two c-quarks have been discovered. These states are expected in QCD and therefore provide a test for the theory. There should be double beauty baryons, and also an intriguing possibility that baryons ${\bf B}_{bc}$ with a c-quark, a b-quark. These states are yet to be discovered. The main decay modes of ${\bf B}_{bc}$  are expected to be weak processes from theoretical understanding of their mass spectrum. These decay modes can provide crucial information about these heavy baryons ${\bf B}_{bc}$. We analyze two body hadronic weak decays for ${\bf B}_{bc}$ using $SU(3)$ flavor symmetry. Any one of the $c$ and $b$ decays will induce ${\bf B}_{bc}$ to decay. We find that the Cabibbo allowed decays ${\bf B}_{bc} \to {\bf B}_b + M$ due to $c \to s u \bar d$ can be crucial for exploration.
The LHC may have the sensitivity to discover such decays.
Other ${\bf B}_{bc}$ decays due to $b \to c q' \bar q$ are sub-leading.
Several relations among branching ratios are obtained which can be used to test
$SU(3)$ flavor symmetry.
\end{abstract}

\maketitle

\section{Effective Hamiltonian of weak ${\bf B}_{bc}$ decays}

Heavy b and c quarks can be baryon constituents as long as they form color singlets with other constituents inside the baryon according to QCD. Baryons with one c-quark and also baryons with one b-quark have been found a long time ago~\cite{pdg}. Two years ago, a baryon with two  c-quarks, the $\Xi_{cc}^{++}$, had also been discovered~\cite{Aaij:2017ueg}. They expected baryons with two b-quarks,
or a c-quark and a b-quark yet await to be discovered
experimentally~\cite{Zhang:2011hi,Li:2017ndo,Shi:2017dto,Xing:2018lre}.
With LHCb continuing to collect more and more data, it is hoped that these states will eventually be discovered.

Although there are no experimental data on ground states of ${\bf B}_{bc}$ baryons with constituent quarks $(bcu,\; bcd,\;bcs)$, there are theoretical estimates on their masses. The masses of these baryons are all estimated to be
below the thresholds of strong decays (less than $\sim$ 7.3 GeV~\cite{Roberts:2007ni,Aliev:2012ru}), the mass splittings between them are also below the possible strong decay of a heavier ground state ${\bf B}_{bc}$ decaying into a lighter one plus a pion. There may be some radiative decays from heavier to lighter states. The lowest state will decay through weak interactions. After being produced, the identification of ${\bf B}_{bc}$ ground states will be through their dominant weak decays. These weak decay modes can provide crucial information about these heavy baryons ${\bf B}_{bc}$. In this work we will analyze two body weak hadronic decays to obtain the main decay modes and also to find relations among different decays as tests for $SU(3)$ flavor symmetry where $u$, $d$ and $s$ form a fundamental representation $3$. $SU(3)$ flavor symmetry has been used to study hadronic decays for hadrons with light quarks and with heavy quarks~\cite{He:2000ys,Fu:2003fy,He:2015fsa,Hsiao:2015iiu,Lu:2016ogy,He:2018joe,Wang:2018utj,Geng:2017mxn,Savage:1989qr}. In the lack of results from established method of first principle calculation from QCD, such as lattice calculation, $SU(3)$ studies have given many interesting results for hadrons that contain a heavy $b$ quark, and in many cases are quantitatively in agreement with experimental data. One expects that $SU(3)$ studies may also provide some useful information for understanding ${\bf B}_{bc}$ properties and for experimental search.

There are three ${\bf B}_{bc}$ baryons,
\begin{eqnarray}
{\bf B}_{bc}({\bf B}_{bc\;i}) = (bcu,\;bcd,\;bcs) = (\Xi_{bc}^+,\;\Xi_{bc}^0,\;\Omega_{bc}^0)\,.
\end{eqnarray}
These three states form a fundamental representation $3$ of $SU(3)$. The subscript ``i" taking the values 1,2,3 in ${\bf B}_{bc\;i}$ is the $SU(3)$ representation index. Changing it to a superscript, it becomes the anti-representation index. The subscript $bc$ is related to the naming of the particle. We will use the same notions for other particles and also Hamiltonian in our later discussions.

The weak decays  of ${\bf B}_{bc}$ baryons can be induced by the constituent b-quark or c-quark decays.
The dominant c-quark decay is induced by $c \to u \bar q' q$ with $q'$ and $q$ taking the values $d$ and $s$.
These decays are proportional to $\lambda_{q'q}=V_{uq'} V^*_{cq}$ with Wilson coefficients  of order one. These decays are classified as Cabibbo allowed ($\lambda_{ds}=V_{ud} V^*_{cs}$), Cabibbo suppressed ($\lambda_{dd}=V_{ud} V^*_{cd}$ or $\lambda_{ss}=V_{us} V^*_{cs}$),  and doubly Cabibbo suppressed ($\lambda_{sd}=V_{us} V^*_{cd}$) decays.  The lifetimes of the three ${\bf B}_{bc}$ states are estimated
to be an order of a few times $10^{-13}$~s~\cite{Karliner:2014gca,Berezhnoy:2018bde,Cheng:2019sxr}.

The $b$ can decay in several ways at tree level, a) $b\to c \bar c q$, b) $b\to c \bar u q$, c) $b \to u \bar c q$, d) $b\to u \bar u q$.
Among them a) and b) are dominating ones which are proportional to $\lambda^c_{cq} = V_{cb}V^*_{cq}$ and $\lambda^c_{uq}=V_{cb} V_{uq}^*$ with the Wilson coefficients of order one.   c) and d) are proportional to $\lambda^c_{bq}=V_{ub}V_{cq}^*$ and $\lambda^u_{bq}=V_{ub} V_{uq}^*$ with Wilson coefficients of order one. At one loop level, there are also the penguin decay modes $b \to q \sum_{q'=u,d,s,c}\bar q' q'$.
These decays are, however, suppressed by loop induced small Wilson coefficients. In our later discussions we will only keep decays with q=d for a) and b), respectively,
and neglect the suppressed parts~\cite{Zhang:2018llc}.
Note that even the two largest classes of $b$-quark induced decays are suppressed by a factor of
{$|V_{cb}/V_{cs}|^2 \sim 1.7\times 10^{-3}$ compared with Cabibbo allowed c-quark decays. But they are different from c-quark decay induced processes and we hope experiments may find some different favorable search strategies. For instance,
the $b$-quark induced ${\bf B}_{bc}$ decays
can have a displaced baryon for subsequent decays,
which is able to travel a $10^{3}$ longer distance before decaying
than that from the $c$-quark decay induced ones.
One can hence consider a displaced vertex as a signal of the $b$-decay mode
since the baryon's decay vertex is displaced from the prompt production vertex, which has been
used as a technique of searching for doubly-bottom hadrons~\cite{displaced}.

For the main ${\bf B}_{bc}$ baryon weak decays described above,
the effective Hamiltonians are given by~\cite{Buras:1998raa}
\begin{eqnarray}\label{Heff}
{\cal H}_{eff}^c&=&\frac{G_F}{\sqrt 2}\lambda_{q'q}
\bigg[c_1(\bar u q' )(\bar q c)
+c_2(\bar u_\beta q'_{ \alpha} )(\bar q_\alpha c_\beta)\bigg]\,,\nonumber\\
{\cal H}_{eff}^b&=&\frac{G_F}{\sqrt 2}\lambda^c_{aq}
\bigg[c_1(\bar q a)(\bar c b)
+c_2(\bar q_\beta a_\alpha)(\bar c_\alpha b_\beta)\bigg]\,,
\end{eqnarray}
for the $c$ and $b$-quark decays, respectively.
$G_F$ is the Fermi constant.
$a$ takes the values $c$ and $u$.
In the above $(\bar q_1 q_2)=\bar q_1\gamma_\mu(1-\gamma_5)q_2$,
and the subscripts $(\alpha,\beta)$ denote the color indices.
The Wilson coefficients $c_i$ are scale ($\mu$)-dependent,
with $\mu=m_{b(c)}$ for the $b(c)$ decays.

We now specify the notation of $SU(3)$ group tensor properties for the effective Hamiltonians, omitting Lorentz structure,
for  $c\to q_j\bar q_i q_k$ and $b\to c \bar q_i q_j$. We will denote them by $H^{jk}_i$ for $H^c_{eff}$,
$H^i_j$ for $H^b_{eff}$ with $a= u$, and $H^i$ for $H^b_{eff}$ with $a=c$, respectively. Their nonzero entries are given by~\cite{He:2018joe,Hsiao:2020iwc}
\begin{eqnarray}\label{Hijk}
&&H^{31}_2=\lambda_{ds},\;\;\;\; H^{21}_2=\lambda_{dd},
\;\;\;\;H^{31}_3=\lambda_{ss}, \;\;\;\;H^{21}_3=\lambda_{sd}\, \nonumber\\
&&H^2_1=\lambda^c_{ud}\,,\;\;\;\;H^3_1=\lambda^c_{us}\,,\;\;\;\;
H^2 = \lambda^c_{cd}\,,\;\;\;\;H^3 = \lambda^c_{cs}\;.
\end{eqnarray}
$H^{jk}_i$ contains a $\bar 3_H$, a $6_H$ and a $\overline{15}_H$~\cite{Savage:1989qr}.
Here the subscript for the representations indicates where they come from, and the $H$ shows that the given representation comes from the effective Hamiltonian.
When calculating various decays, the results for $\bar 3_H$
will be proportional to $\lambda_{dd} + \lambda_{ss}$. Using the unitarity property of the CKM matrix,
this combination is equal to $- V_{ub}V^*_{cb}$
which is much smaller than any of the $\lambda_{q'q}$ and can be safely neglected. In other words,
$\lambda_{ss} = -\lambda_{dd}$ will be a good approximation for our purpose.
$H^i_j$  and $H^i$ transform as a $8_H$ and $\bar 3_H$, respectively.
The above effective Hamiltonians can induce various hadronic ${\bf B}_{bc}$ decays,
including two and multi-body channels~\cite{Shi:2017dto}.
We will concentrate on the two-body hadronic decays
which may provide the most promising chances for experimental measurements.

\section{The two body ${\bf B}_{bc}$ decay modes}

We now list some of the interesting two body decay modes of ${\bf B}_{bc}$. The effective Hamiltonian $H^{jk}_i$ can induce ${\bf B}_{bc}$ to decay into an octet-$8_{\bf B}$ (decuplet-$10_{{\bf B}'}$) baryon ${\bf B}\;({\bf B}')$
and a $\bar 3_{M_b}$ b-meson $M_b$
and also into a $\bar 3_{{\bf B}_b}$ ($6_{{\bf B}'_b}$) b-baryon
${\bf B}_b\;({\bf B}'_b)$ plus an octet-$8_M$ meson $M$,
\begin{eqnarray}
{\bf B}_{bc} \to {\bf B}^{(\prime)} + M_b,\;\;\;\;
{\bf B}_{bc} \to {\bf B}_b^{(\prime)}+ M\,,
\end{eqnarray}
where ${\bf B}^{(\prime)}$ stands for the octet (decuplet) baryon and
${\bf B}_Q^{(\prime)}$ the anti-triplet (sextet) $Q$-baryon with $Q=(b,c)$.

$H^i_j$ can induce ${\bf B}_{bc}$ to decay into
a triplet-$3_{{\bf B}_{cc}}$ double charmed baryon ${\bf B}_{cc}$ and a meson $M$,
and also into a $\bar 3_{{\bf B}_c}$ ($6_{{\bf B}'_c}$) c-baryon
plus $\bar 3_{M_c}$ $c$-meson $M_c$
\begin{eqnarray}
{\bf B}_{bc} \to {\bf B}_{cc} + M,\;\;\;\;{\bf B}_{bc} \to {\bf B}^{(\prime)}_c + M_c\,.
\end{eqnarray}

$H^i$ can induce the following decay modes
\begin{eqnarray}
{\bf B}_{bc} \to {\bf B}_{cc} +M_{\bar c},\;\;\;\;
{\bf B}_{bc} \to {\bf B}_c^{(\prime)}+ M_{c\bar c}\;,
\end{eqnarray}
where $M_{\bar c}$ denotes the anti-particle of $M_c$,
and $M_{c\bar c}=(\eta_c,J/\psi)$.

The usual octet-8 ${\bf B}$ baryon has components ${\bf B}^i_j$: ($n,\;p,\;,\Sigma^{\pm,0}, \Xi^{-,0},\;\Lambda$).
The usual decuplet-10 ${\bf B}' $ has components ${\bf B}'_{ijk}$: ($\Delta^{++,\pm, 0}, \;\Sigma^{\prime\; \pm, 0},\;\Xi^{\prime\;-,0},\;
\Omega^-$) with the subscripts $ijk$ being totally symmetric. The usual octet-8 pseudoscalar meson $M$ has components $M^i_j$:
($\pi^{\pm,0},\;K^\pm,\;K^0,\;\bar K^0,\;\eta$). The baryon states containing heavy quarks are indicated by
\begin{eqnarray}\label{B8B10}
&&{\bf B}_{cc} ({\bf B}_{cc\;i})=(\Xi_{cc}^{++},\Xi_{cc}^+,\Omega_{cc}^+), \nonumber\\
&&\nonumber\\
&&{\bf B}_b ({\bf B}_{b\;ij})=\left(\begin{array}{ccc}
0& \Lambda_b^0 & \Xi_b^0\\
-\Lambda_b^0&0&\Xi_b^- \\
-\Xi_b^0&-\Xi_b^-&0
\end{array}\right)\,,\;
{\bf B}'_{b}({\bf B}'_{b\; ij})=\left(\begin{array}{ccc}
\Sigma^{+}_{b}& \frac{1}{\sqrt{2}}\Sigma^{0}_{b} & \frac{1}{\sqrt{2}}\Xi'^{0}_{b}\\
 \frac{1}{\sqrt{2}}\Sigma^{0}_{b} &\Sigma^-_b  & \frac{1}{\sqrt{2}}\Xi'^{-}_{b}\\
 \frac{1}{\sqrt{2}}\Xi'^{0}_{b} & \frac{1}{\sqrt{2}}\Xi'^{-}_{b} &\Omega^-_b
\end{array}\right)\,,\nonumber\\
&&\nonumber\\
&&{\bf B}_c({\bf B}_{c\;ij})=\left(\begin{array}{ccc}
0& \Lambda_c^+ & \Xi_c^+\\
-\Lambda_c^+&0&\Xi_c^0 \\
-\Xi_c^+&-\Xi_c^0&0
\end{array}\right)\,,\;
{\bf B}'_{c}({\bf B}'_{c\;ij})=\left(\begin{array}{ccc}
\Sigma^{++}_{c}& \frac{1}{\sqrt{2}}\Sigma^{+}_{c} & \frac{1}{\sqrt{2}}\Xi'^{+}_{c}\\
 \frac{1}{\sqrt{2}}\Sigma^{+}_{c} &\Sigma^0_c  & \frac{1}{\sqrt{2}}\Xi'^{0}_{c}\\
 \frac{1}{\sqrt{2}}\Xi'^{+}_{c} & \frac{1}{\sqrt{2}}\Xi'^{0}_{c} &\Omega^0_c
\end{array}\right)\,,
\end{eqnarray}
and the component meson states in $M_{b,c}$ read
\begin{eqnarray}\label{Mbc}
M_b(M^i_b)&=&(B^-,\bar B^0,\bar B^0_s)\,,\;\;\;\;M_c(M^i_c) = (D^0,D^+,D^+_s)\,.
\end{eqnarray}
The octet baryon can also be written as
${\bf B}_{ijk} = \epsilon_{ijl} {\bf B}^l_k$.
This form will be particularly useful in drawing topological diagrams for the decay amplitude at quark level.

\section{SU(3) Invariant Amplitudes}
We now provide some details for the $SU(3)$ invariant decay amplitudes for the decays mentioned in previous sections.
To obtain $SU(3)$ amplitudes, one just needs to contract all upper and lower indices of the hadrons and the Hamiltonian to form
all possible $SU(3)$ singlets and associate each with a parameter which lumps up the Wilson coefficients and unknown hadronization effects.
These parameters can be determined theoretically and experimentally. Our emphasis will not be on how to determine these hadronic parameters, but to identify the dominant modes and some relations for experimental search. We will normalize the decay amplitudes as ${\cal A}\equiv (G_F/\sqrt{2}){\cal M}$
and leave the spinor and Lorentz structure out.

The $H^{jk}_i$ induced ${\bf B}_{bc}$ decays are given by
\begin{eqnarray}
{\bf B}_{bc} \to {\bf B}^{(\prime)}+ M_b,\;\;\;\;
{\bf B}_{bc} \to {\bf B}_b^{(\prime)} + M\,.
\end{eqnarray}
For ${\bf B}_{bc}\to{\bf B}^{(\prime)}M_b$,
we have
\begin{eqnarray}\label{BbcBMb}
{\cal M}({\bf B}_{bc}\to {\bf B}M_b)&=&
d_1{\bf B}_{bc}^i H^{jk}_i {\bf B}_{ljk}M_b^l+
d_2 {\bf B}_{bc}^i H^{jk}_i {\bf B}_{jkl} M_b^l + d'{\bf B}_{bc}^i H^{jk}_i {\bf B}_{klj} M_b^l \nonumber\\
&+&d_3 {\bf B}_{bc}^i H^{jk}_l {\bf B}_{ijk}M_b^l+
d_4{\bf B}_{bc}^i H^{jk}_l {\bf B}_{jki}M_b^l + d''{\bf B}_{bc}^i H^{jk}_l {\bf B}_{kij}M_b^l \,,
\nonumber\\
{\cal M}({\bf B}_{bc}\to {\bf B}^\prime M_b)&=&
d_1^{\prime}{\bf B}_{bc}^i H^{jk}_i {\bf B}^{\prime}_{ljk}M_b^l+
d_2^{\prime}{\bf B}_{bc}^i H^{jk}_l {\bf B}^{\prime}_{ijk}M_b^l\,.
\end{eqnarray}
We show the corresponding topological diagrams for these two classes of decays in Fig.~\ref{fig1}.

In the above, we have neglected terms needing to contract two indices of the Hamiltonian
$H^{ij}_i$.
Because this will result in the combination of $\lambda_{dd} + \lambda_{ss}$,
leading to $-V_{ub}V^*_{cb}$, whose absolute value $\sim 10^{-4}$
is very small compared to $|\lambda_{dd,ss}|\simeq 0.22$.

One needs to make sure if the above amplitudes are all independent.
This can be checked by group theoretical considerations. The number of
independent amplitudes for ${\bf B}_{bc} \to {\bf B}^{(\prime)}+ M_b$ is equivalent to the number of
$SU(3)$ singlets in the production of $\bar 3_{{\bf B}_{bc}}\times (\bar 3_H, 6_H, \overline{15}_H) \times 8_{\bf B} (10_{{\bf B}'}) \times \bar 3_{M_b}$.
With  $\bar 3_H$ for ${\bf B}_{bc} \to {\bf B} + M_b$, 2 singlets can be formed which correspond to 2 invariant amplitudes. However, to get a
$\bar 3$ requires to contract two indices and to have $H^{ij}_i$ whose contributions are small, proportional to $\lambda_{dd}+\lambda_{ss}$ and can be neglected.
For $6_H$ and $\overline{15}_H$, each produces 2 singlets.
Therefore there are total 4 independent invariant amplitudes.
Naively there are 6 terms as can be seen in the above equation. However
using the identity $B_{ijk} + B_{kij} + B_{jki} = 0$, one can rewrite $d' B_{klj} = - d' B_{ljk} - d' B_{jkl}$.
Then $d'$ term can be absorbed by replacing $d_1-d'$ and $d_2-d'$ by the $d_1$ and $d_2$ terms, respectively.
Similarly, $d''$  can be absorbed into $d_3$ and $d_4$ terms.
We will choose $d_1,\;d_2,\;d_3,\;d_4$ as independent invariant amplitudes in our later discussions.
For ${\bf B}_{bc} \to {\bf B}' + M_b$, neglecting 1 invariant amplitude from  $\bar 3_H$ for the same reason as before, there are 2 independent amplitudes with $6_H$ and $\overline{15}_H$ contributing to 0 and 2, respectively. We use the notations in the second equation of eq.~(\ref{BbcBMb}).

%
\begin{figure}
\setlength{\abovecaptionskip}{-10pt}
\centering
\includegraphics[width=13cm,angle=0]{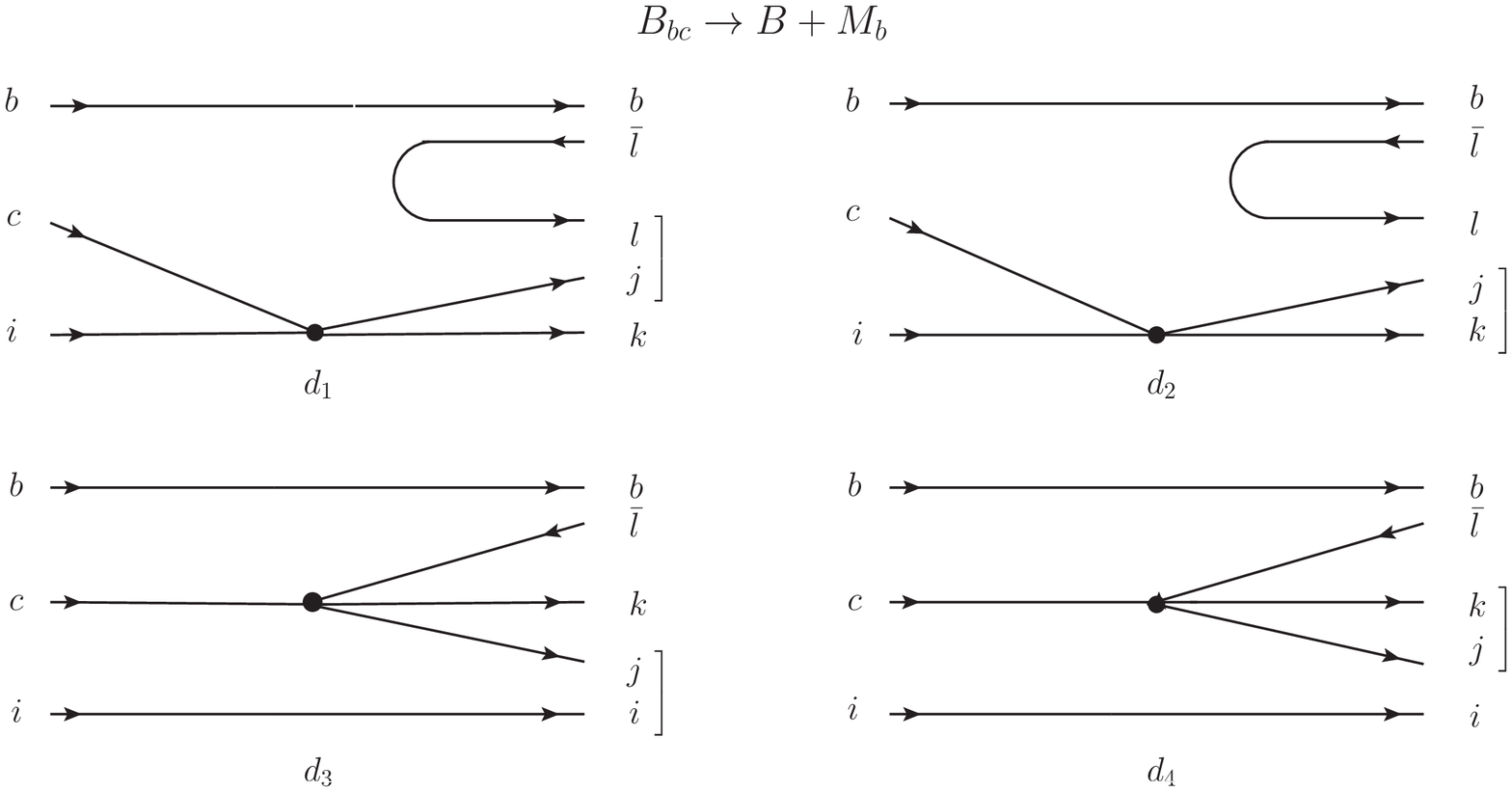}
\\
\vspace{1cm}
\includegraphics[width=13cm,angle=0]{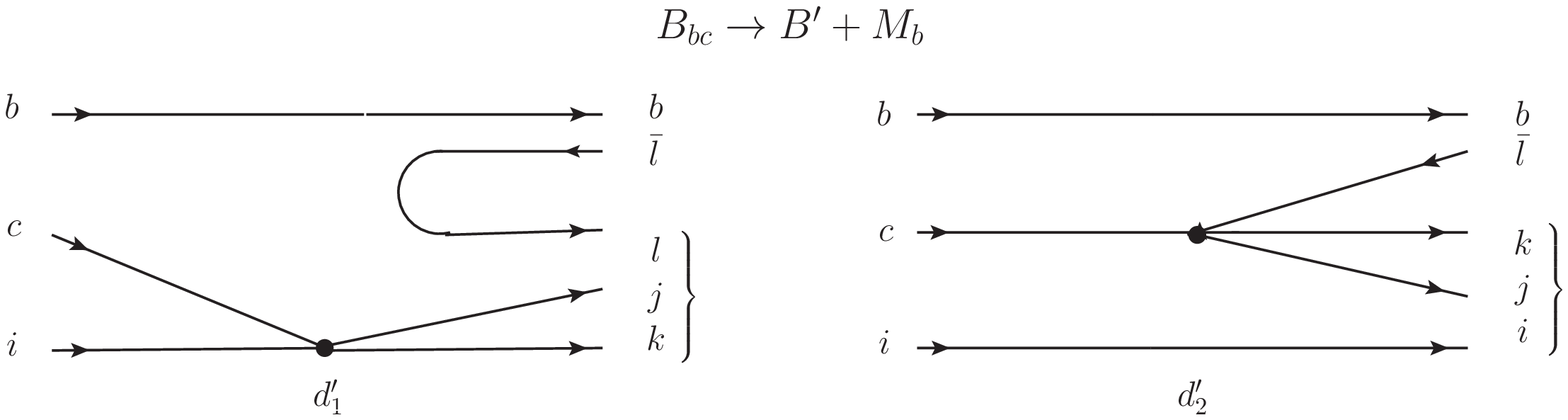}
\vspace{2cm}

\caption{Topological diagrams for ${\bf B}_{bc}\to {\bf B}^{(\prime)}M_b$.
The two quarks next to ``]'' are anti-symmetric, and  the three quarks next to ``\}'' are totally symmetric.}
\label{fig1}
\end{figure}
%

For ${\bf B}_{bc} \to {\bf B}_b^{(\prime)}+ M$, we can naively write down the following terms
\begin{eqnarray}
{\cal M}({\bf B}_{bc}\to {\bf B}_bM)&=&
e_1{\bf B}_{bc}^i H^{jk}_i {\bf B}_{b\;jl}M^l_k+
e_2 {\bf B}_{bc}^i H^{jk}_i {\bf B}_{b\;kl} M^l_j
+e_3{\bf B}_{bc}^i H^{jk}_l {\bf B}_{b\;jk}M^l_i\nonumber\\
&+&e_4{\bf B}_{bc}^i H^{jk}_l {\bf B}_{b\;ik}M^l_j
+ e_5{\bf B}_{bc}^i H^{jk}_l{\bf B}_{b\;ij} M^l_k\,,
\nonumber\\
{\cal M}({\bf B}_{bc}\to {\bf B}_b^\prime M)&=&
e'_1{\bf B}_{bc}^i H^{jk}_i {\bf B}_{b\;jl}^\prime M^l_k+
e'_2{\bf B}_{bc}^i H^{jk}_i {\bf B}_{b\;kl}^\prime M^l_j+
e'_3{\bf B}_{bc}^i H^{jk}_l {\bf B}_{b\;jk}^\prime M^l_i
\nonumber\\
&+&
e'_4{\bf B}_{bc}^i H^{jk}_l {\bf B}_{b\;ik}^\prime M^l_j+
e'_5{\bf B}_{bc}^i H^{jk}_l{\bf B}_{b\;ij}^\prime M^l_k \,.
\end{eqnarray}
The corresponding topological diagrams for these decays are shown in Fig.~\ref{fig2}.

The ${\bf B}_{bc} \to {\bf B}_b + M$ decay has the same group structure as ${\bf B}_{bc} \to {\bf B} + M_b$,
therefore this class of decay has only 4 independent amplitudes.
One of the parameters can be absorbed into others.
For example,
with the appearances of $(e_1-e_5)$, $(e_2+e_5)$, $(e_3-e_5)$ and $(e_4+e_5)$
in the full expansion of ${\cal M}({\bf B}_{bc}\to {\bf B}_bM)$,
the $e_5$ term is redundant. We choose to work with the convention with $e_5=0$.}
But for ${\bf B}_{bc} \to {\bf B}'_b + M$, neglecting $\bar 3_H$ contribution, there are total 5 independent invariant amplitudes with $6_H$ and $\overline{15}_H$ contributing to 2 and 3, respectively. We use the above terms associated with  $e'_i$.

%

\begin{figure}
\centering
\includegraphics[width=13cm,angle=0]{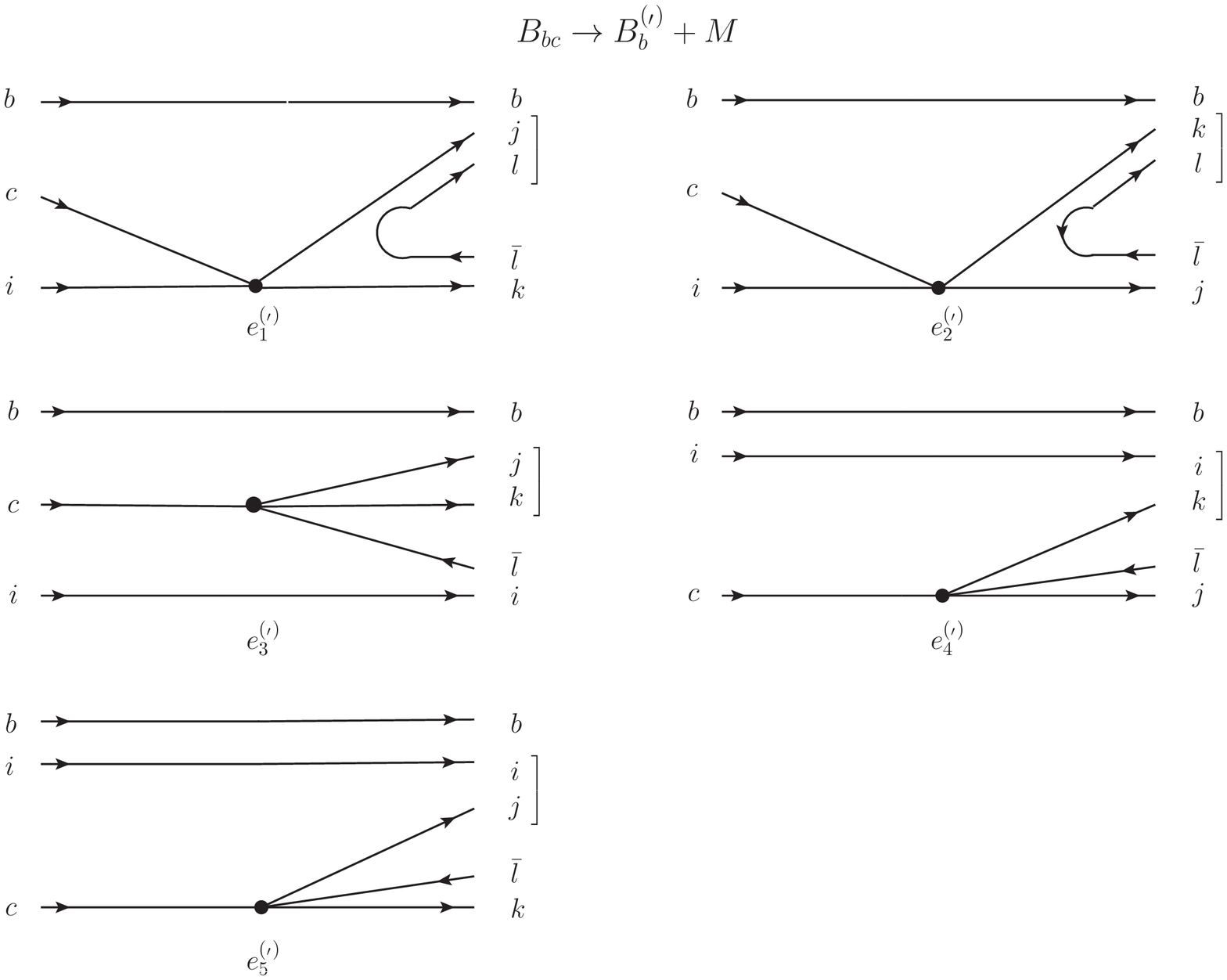}
\vspace{1cm}
\caption{Topological diagrams for ${\bf B}_{bc}\to {\bf B}^{(\prime)}_b M$.
The two quarks next to  $``]"$ are anti-symmetric (symmetric) for ${\bf B}_b$ $(\bf B^{\prime}_b)$, respectively.
For ${\bf B}_{bc} \to {\bf B}_b+ M$, one should remove the $e_5$ term.}
\label{fig2}
\end{figure}
%

\begin{figure}
\setlength{\abovecaptionskip}{-30pt}
\centering
\includegraphics[width=13cm,angle=0]{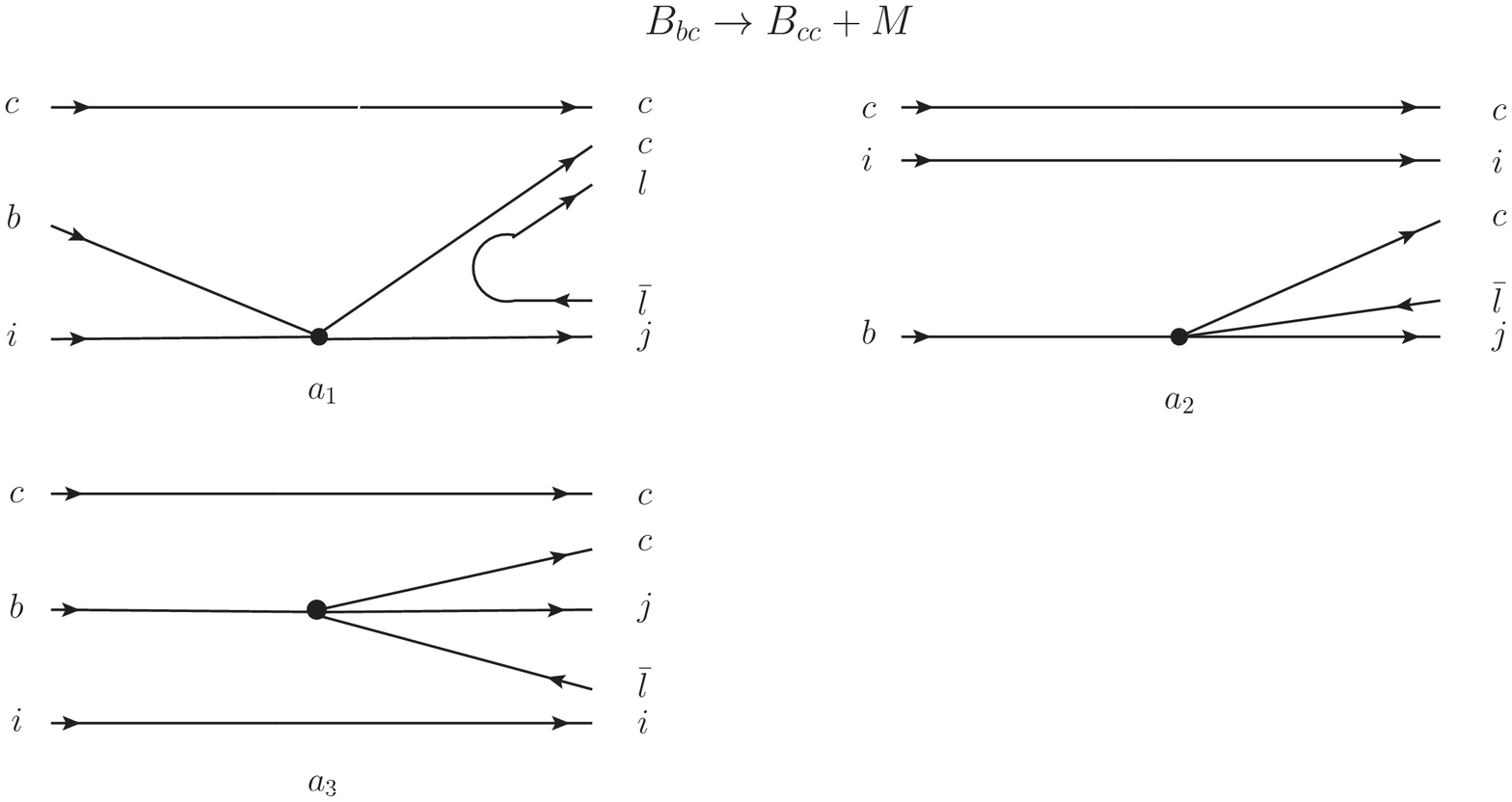}
\\
\vspace{1cm}
\includegraphics[width=13cm,angle=0]{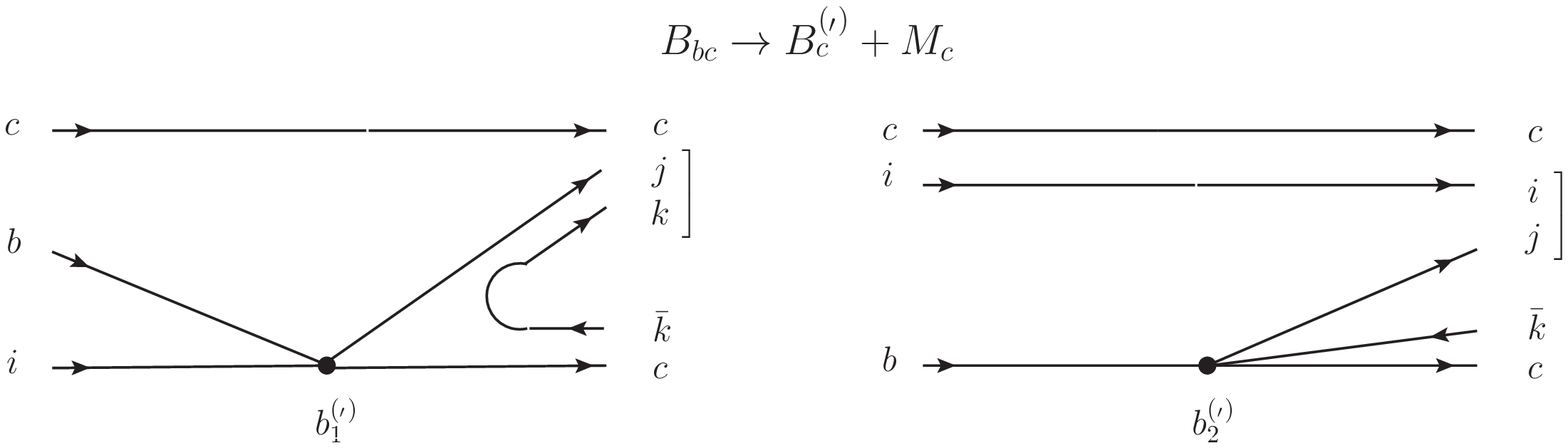}
\vspace{2cm}
\caption{Topological diagrams for
${\bf B}_{bc}\to {\bf B}_{cc} M$ (top panel) and ${\bf B}_{bc}\to {\bf B}^{(\prime)}_c M_c$ (bottom panel),
where the two quarks next to $``]"$ are anti-symmetric (symmetric) for ${\bf B}_c$ $({\bf B}_c^\prime)$.}
\label{fig3}
\end{figure}

For $H^i_j$ induced ${\bf B}_{bc}$ decays, ${\bf B}_{bc} \to {\bf B}_{cc} + M$ and ${\bf B}_{bc} \to {\bf B}_c^{(\prime)}+ M_c$,
we have
\begin{eqnarray}
&&{\cal M}({\bf B}_{bc}\to {\bf B}_{cc}M) =
a_1 {\bf B}_{bc}^i H^j_i {\bf B}_{cc\;l} M^l_j+a_2 {\bf B}_{bc}^i H^j_l {\bf B}_{cc\;i} M^l_j+
a_3 {\bf B}_{bc}^i H^j_l {\bf B}_{cc\;j} M^l_i\,,\nonumber\\
&&{\cal M}({\bf B}_{bc}\to {\bf B}^{(\prime)}_{c}M_c)=
b_1^{(\prime)}{\bf B}_{bc}^i H^j_i {\bf B}_{c\;jk} M_c^k+
b_2^{(\prime)}{\bf B}_{bc}^i H^j_k {\bf B}_{c\;ji} M_c^k\,.
\end{eqnarray}
The corresponding topological diagrams are shown in Fig.~\ref{fig3}.
Similar group theoretical analysis shows that the above invariant amplitudes are all independent ones.

%

\begin{figure}
\setlength{\abovecaptionskip}{-20pt}
\centering
\includegraphics[width=13cm,angle=0]{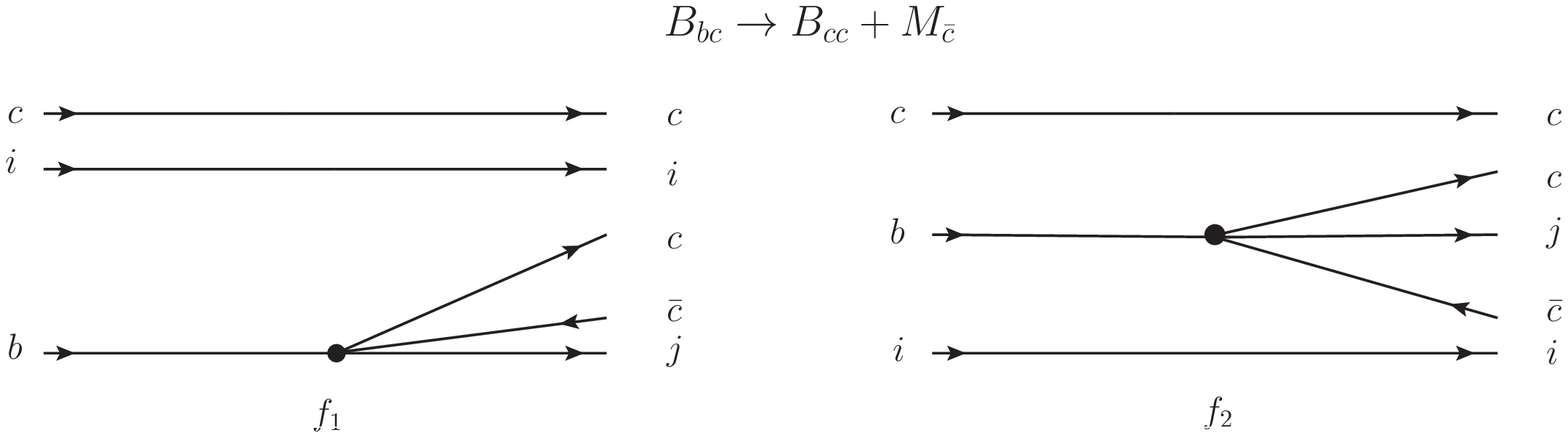}
\\
\vspace{1cm}
\includegraphics[width=6cm]{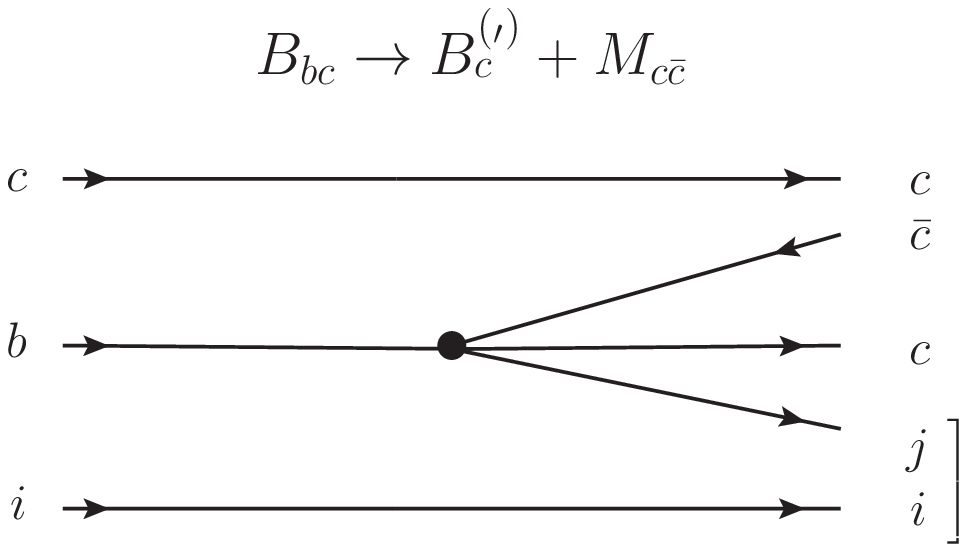}
\vspace{2cm}
\caption{Topological diagrams for
${\bf B}_{bc}\to {\bf B}_{cc} M_{\bar c}$ (top panel) and
${\bf B}_{bc}\to {\bf B}^{(\prime)}_c M_{c\bar c}$ (bottom panel).
The two quarks next to ``]'' are anti-symmetric (symmetric) for ${\bf B}_c$ $(\bf B^\prime_c)$.}
\label{fig4}
\end{figure}

Finally, for $H^i$ induced ${\bf B}_{bc}$ decays,
${\bf B}_{bc} \to {\bf B}_{cc} + M_{\bar c}$ and
${\bf B}_{bc} \to {\bf B}_c^{(\prime)} + M_{c\bar c}$, we have
\begin{eqnarray}
&&{\cal M}({\bf B}_{bc}\to {\bf B}_{cc}M_{\bar c}) =
f_1 {\bf B}_{bc}^i H^j {\bf B}_{cc\;i} M_{\bar c\;j}+
f_2 {\bf B}_{bc}^i H^j{\bf B}_{cc\;j} M_{\bar c\;i}\;,\nonumber\\
&&{\cal M}({\bf B}_{bc}\to {\bf B}^{(\prime)}_c M_{c\bar c})=
g^{(\prime)}_{c\bar c} {\bf B}_{bc}^i H^j {\bf B}^{(\prime )}_{c\;ij} M_{c\bar c} \,.
\end{eqnarray}
The corresponding topological diagrams are shown in Fig.~\ref{fig4}.
%

\section{Decay modes for experimental analysis}

For experimental discovery of ${\bf B}_{bc}$, the most favored decay modes will certainly be those with
large branching ratios, and at the same time particles in the final state can
be easily identified and analyzed. For the decay modes discussed in previous sections, there are two
classes  of decays, one is decay induced by $c$ decays and another induced by $b$ decays. We discuss $c$ induced decays below,
\begin{eqnarray}
{\bf B}_{bc} \to {\bf B}^{(\prime)}+ M_b,\;\;\;\;
{\bf B}_{bc} \to {\bf B}_b^{(\prime)} + M\,.
\end{eqnarray}

Among ${\bf B}_{bc} \to {\bf B}^{(\prime)}+ M_b$ and ${\bf B}_{bc} \to {\bf B}_b^{(\prime)}+ M$, the first one has advantages compared with the second one
because the final states ${\bf B}^{(\prime)}$ and $M_b$ are all well studied experimentally. The further decay branching ratios for many of them are known to good precisions. For the second one although properties of $M$ are well known,  the final baryon ${\bf B}_b^{(\prime)}$ properties are not as well known as ${\bf B}^{(\prime)}$ baryons and therefore this class of decays may not be as easy as ${\bf B}_{bc} \to {\bf B}^{(\prime)}+ M_b$ for analysis.
But still many decays modes of  ${\bf B}_b^{(\prime)}$  have been measured.
The second one can also serve to confirm the discovery of ${\bf B}_{bc}$ and study its detailed properties and may even luckily become the discovered modes if identification of ${\bf B}_b^{(\prime)}$ can be optimized.

From Tables~\ref{tab1} and \ref{tab2}, we can identify the following Cabibbo allowed decays as
\begin{eqnarray}
&\mbox{$B_{bc} \to B + M_b$ type}:&\;\;\Xi_{bc}^0 \to \Xi^0 \bar B^0_s,\;\;\Omega_{bc}^0 \to \Xi^0 \bar B^0,\;\;
\Xi_{bc}^0 \to \Sigma^+ B^-,\nonumber\\
&&\;\;\Xi_{bc}^+ \to \Sigma^+ \bar B^0,\;\;\Xi_{bc}^0 \to \Lambda \bar B^0,\;\;\Xi_{bc}^0 \to \Sigma^0 \bar B^0.\nonumber\\
&\mbox{$B_{bc} \to B^\prime + M_b$ type}:
&\;\;\Xi_{bc}^0 \to \Sigma^{\prime+} B^-,\;\;\Xi_{bc}^0 \to \Xi^{\prime0} \bar B^0_s,\;\;\Omega_{bc}^0 \to \Xi^{\prime0} \bar B^0,\nonumber\\
&&\;\;\Xi_{bc}^+ \to \Sigma^{\prime+} \bar B^0,\;\;\Xi_{bc}^0 \to \Sigma^{\prime0} \bar B^0.\label{Mbtype}\\
&\mbox{$B_{bc} \to B_b + M$ type}: &\;\;\Xi_{bc}^0 \to \Xi_b^-\pi^+,\;\;\Xi_{bc}^+ \to \Xi_b^0 \pi^+,
\;\;\Xi_{bc}^0 \to \Xi_b^0\pi^0,\nonumber\\
&&\;\;\Omega_{bc}^0 \to \Xi_b^0 \bar K^0,\;\;\Xi_{bc}^0 \to \Xi_b^0\eta,
\;\;\Xi_{bc}^0\to\Lambda_b\bar K^0.\nonumber\\
&\mbox{$B_{bc} \to B^\prime_b + M$ type}:&\;\;
\Xi_{bc}^0 \to \Omega_b^{-} K^+,\;\;\Xi_{bc}^0 \to \Sigma_b^{+}K^-,\;\;\Xi_{bc}^0 \to \Xi_b^{\prime0} \eta,\nonumber\\
&&\;\;\Xi_{bc}^+ \to \Sigma_b^{+}\bar K^0,\;\;\Omega_{bc}^0 \to \Omega_b^- \pi^+,
\;\;\Xi_{bc}^+ \to \Xi_b^{\prime0} \pi^+,\nonumber\\
&&\;\;\Xi_{bc}^0 \to \Xi_b^{\prime-} \pi^+,\;\;\Xi_{bc}^0 \to \Sigma_b^{0}\bar K^0,
\;\;\Omega_{bc}^0 \to \Xi_b^{\prime0} \bar K^0,\;\;\Xi_{bc}^0 \to \Xi_b^{\prime0} \pi^0.\label{Mtype}
\end{eqnarray}

We expect the above decay modes to be the dominant ones which are some of the most likely to-be-discovered decay modes.
Since the lifetime of ${\bf B}_{bc}$ is mainly determined by $c\to s u \bar d$, the lifetime of $\Xi^+_{bc}$
would be similar to $\Xi_{cc}^{++}$ and the decay phase spaces and the particle masses are different and therefore
may deviate from each other. This expectation is in agreement with some theoretical estimates.
For example, in Ref.~\cite{Cheng:2019sxr} it is estimated to be a few times of $10^{-13}$~s.
A similar argument would lead to the expectation that the decay width of $\Xi^{+}_{bc} \to \Xi^0_b + \pi^+$
is similar to $\Xi^{++}_{cc} \to \Xi^+_c + \pi^+$ up to phase space modifications.
Taking factorizable contributions as the example for an order of magnitude estimate,
we derive that ${\cal M}(\Xi_{Qc}\to \Xi_Q\pi^+)=
i\lambda_{ds}a_1 f_\pi q^\mu \langle \Xi_Q|(\bar sc)|\Xi_{Qc}\rangle$,
where
the decay constant  $f_\pi$ is from $\langle \pi^+|(\bar ud)|0\rangle=if_\pi q^\mu$,
and $a_1=c_1+c_2/N_c$ with $N_c$ the color number.
With $\langle \Xi_Q|(\bar sc)|\Xi_{Qc}\rangle\simeq \bar u_{\Xi_Q}
(f_1\gamma_\mu-g_1\gamma_\mu\gamma_5)u_{\Xi_{Qc}}$~\cite{Cheng:2020wmk},
we obtain
\begin{eqnarray}
\Gamma(\Xi_{Qc}\to \Xi_Q \pi^+)=\frac{G_F^2}{32\pi}(\lambda_{ds}a_1 f_\pi)^2
m_{\Xi_{Qc}}^3\bigg(1-\frac{m_{\Xi_Q}^2}{m_{\Xi_{Qc}}^2  }\bigg)^3(f_1^2+g_1^2)\,,
\end{eqnarray}
where $m_\pi\simeq 0$ has been used.
We can hence have
$\Gamma(\Xi^+_{bc} \to \Xi_b^0 + \pi^+)/\Gamma(\Xi^{++}_{cc} \to \Xi_c^+ + \pi^+)
\sim (m_{\Xi^+_{bc}}/m_{\Xi^{++}_{cc} })^3
[(1-m^2_{\Xi_b^0}/m^2_{\Xi^+_{bc}})/(1-m^2_{\Xi_c^+}/m^2_{\Xi^{++}_{cc}})]^3
\approx 1.4$ with
$(m_{\Xi^+_{bc}}/m_{\Xi^{++}_{cc}},m_{\Xi_b^0}/m_{\Xi^+_{bc}},m_{\Xi_c^+}/m_{\Xi^{++}_{cc}})
\simeq (1.93,0.83,0.68)$~\cite{pdg,Roberts:2007ni}.
This leads to ${\cal B}(\Xi^+_{bc} \to \Xi_b^0 + \pi^+)$
an order of a few times $10^{-2}$~\cite{Cheng:2020wmk},
in agreement with the calculations in Refs.~\cite{Wang:2017mqp,Zhao:2018mrg,
Shi:2019hbf,Gerasimov:2019jwp}. One expects other Cabibbo-allowed
${\bf B}_{bc} \to {\bf B}_b^{(\prime)} + M$ and also ${\bf B}_{bc} \to {\bf B}^{(\prime)} +M_b$
branching fractions to have similar order of magnitudes~\cite{Wang:2017mqp,Zhao:2018mrg,
Shi:2019hbf,Gerasimov:2019jwp}.

According to the gluon-gluon fusion mechanism
that dominantly produces the baryons with two heavy quarks~\cite{Zhang:2011hi,
Berezhnoy:1998aa,Kiselev:2001fw}, the cross section ($X$) for the $\Xi_{bc}$ production is estimated
to be $(17\pm 3)$ and $(35\pm 7)$~nb
at LHC for the center-of-mass (c.m.) energy $\sqrt S=7$~GeV and $\sqrt S=14$~GeV,
respectively, where the theoretical errors mainly consider the uncertainties from
the quark masses $m_b$ and $m_c$, together with
the non-perturbative effects and the factorization scale.
Since the LHC luminosity ($L$) of the c.m. energy $\sqrt S=(7,14)$~GeV
can be (10,100)~fb$^{-1}/yr$~\cite{Zhang:2011hi}, using $X\cdot L$
we obtain the $\Xi_{bc}$ events about $(1.7\pm 0.3)\times 10^8$ and $(3.5\pm 0.7)\times 10^9$ per year.
Note that the uncertainties of around 20\% come from the cross sections,
which demonstrates the reliability of the theoretical estimates. On the other hand,
to discover $\Xi_{bc}$ and determine its mass,
the decay products have to be fully constructed, involving the branching ratios
of the discovery channels and the subsequent decays.
The most promising discovery channels should be
the Cabibbo-allowed ones with $\bf B$ or ${\bf B}_b$
in Eqs.~(\ref{Mbtype}) and (\ref{Mtype}),
which cause less subsequent decays compared to
those with $\bf B'$ and ${\bf B}'_b$.
Let us take $\Xi_{bc}^0 \to \Lambda+\bar B^0$ and $\Xi^+_{bc} \to \Xi_b^0 + \pi^+$ as our illustrations.
For $\Xi_{bc}^0 \to \Lambda \bar B^0$,
since $\Lambda\to p\pi^-$ and $\bar B^0\to D^+\pi^-$ can be suitable subsequent decays,
with ${\cal B}(\Xi_{bc}^0 \to \Lambda+\bar B^0)\sim 10^{-2}$,
${\cal B}(\Lambda\to p\pi^-)\simeq 64\%$, 
${\cal B}(\bar B^0\to D^+\pi^-)\simeq 2.5\times 10^{-3}$ and 
${\cal B}(D^+\to\pi^+\pi^0)\simeq 1.2\times 10^{-3}$,
we estimate that at least $10^{8}$ $\Xi_{bc}$ events are needed.
For $\Xi^+_{bc} \to \Xi_b^0 + \pi^+$, we have $\Xi_b^0\to \Xi^0 J/\Psi$
as the subsequent decay with ${\cal B}=10^{-4}-10^{-3}$~\cite{Hsiao:2015cda}.
In addition to ${\cal B}(\Xi^0\to \Lambda\pi^0,\Lambda\to p \pi^-)\simeq 64\%$
and ${\cal B}(J/\Psi\to \mu^+\mu^-+e^+ e^-)\simeq 12\%$~\cite{pdg},
it leads to the estimation of the $10^{6-7}$ $\Xi_{bc}^+$ events at least~\cite{Qian}.
There can be other subsequent processes,
such as $\Xi_b\to\Xi_c+(D_s^+,\pi^+)$,
whose needed $\Xi_{bc}^+$ events are of the same orders.
Similarly, we estimate the other most promising discovery channels, and
summarize the needed $\Xi_{bc}$ events in Table~\ref{tab5}.
Therefore,
the LHC may be able to discover ${\bf B}_{bc}$ given their excellent capabilities of identifying
${\bf B}^{(\prime)}$, $M_b$ and also reasonably good identifications for ${\bf B}_b$ baryons.
We strongly urge our experimental colleagues to search for those ${\bf B}_{bc}$ decays.

One also sees from the Tables~\ref{tab1} and \ref{tab2} that there are several relations among decay
amplitudes induced by $c\to u \bar q' q$.
Those Cabibbo allowed decays offer good chances to be tested experimentally,
given by
\begin{eqnarray}\label{re_2}
&&
{\cal M}(\Xi_{bc}^0\to \Sigma^{\prime+} B^-) = {\cal M}(
\Xi_{bc}^0\to \Xi^{\prime0} \bar B^0_s) =\frac{1}{\sqrt 3 }\lambda_{ds}d'_1\,,\nonumber\\
&&
{\cal M}(
\Omega_{bc}^0\to \Xi^{\prime0} \bar B^0)= {\cal M}
(\Xi_{bc}^+\to \Sigma^{\prime+} \bar B^0)=\frac{1}{\sqrt 3} \lambda_{ds} d'_2\,,
\end{eqnarray}
and triangle relations
\begin{eqnarray}
&&{\cal M}(\Xi_{bc}^0\to \Sigma^{\prime+} B^-)
+{\cal M}
(\Xi_{bc}^+\to \Sigma^{\prime+} \bar B^0)
-\sqrt 2{\cal M}(\Xi_{bc}^0\to\Sigma'^0 \bar B^0)=0\,,\nonumber\\
&&{\cal M}(\Xi_{bc}^+\to\Xi_b^0\pi^+)
-{\cal M}(\Xi_{bc}^0\to\Xi_b^-\pi^+)+\sqrt 2 {\cal M}(\Xi_{bc}^0\to\Xi_b^0\pi^0)=0\,,\nonumber\\
&&{\cal M}(\Xi_{bc}^+\to\Sigma_b^+\bar K^0)
+{\cal M}(\Xi_{bc}^0\to\Sigma_b^+ K^-)-\sqrt 2{\cal M}(\Xi_{bc}^0\to\Sigma_b^0 \bar K^0)=0\,.
\end{eqnarray}
There are also some relations between Cabibbo allowed, Cabibbo suppressed and doubly Cabibbo suppressed decay modes which can be read off from Tables~\ref{tab1} and \ref{tab2}.
With more and more data being collected, these relations can also serve further examintations.

For $b$-decay induced decays, we have
\begin{eqnarray}
&&
{\bf B}_{bc} \to {\bf B}_{cc} + M,\;\;\;\;
{\bf B}_{bc} \to {\bf B}_c^{(\prime)}+ M_c\,\nonumber\\
&&
{\bf B}_{bc} \to {\bf B}_{cc} +M_{\bar c},\;\;\;
{\bf B}_{bc} \to {\bf B}_c^{(\prime)} + M_{c\bar c}\;.
\end{eqnarray}
As mentioned before that $b$-decay induced modes are suppressed by a factor of
$|V_{cb}/V_{cs}|\simeq 0.04$
compared with the Cabibbo allowed ones in $c$-decay induced modes.
Nonetheless, their branching fractions are not necessarily small.
Using
the ${\bf B}_{bc}\to {\bf B}_{c},{\bf B}_{cc}$ transition form factors obtained in Refs.~\cite{Hu:2020mxk},
we estimate that
${\cal B}(\Xi_{bc}^+\to\Sigma_c^+ D^0)\sim 10^{-6}$,
${\cal B}(\Xi_{bc}^+\to\Xi_c^+ \eta_c)\sim 10^{-5}$,
${\cal B}(\Xi_{bc}^+\to\Xi_{cc}^{++} \pi^-)\sim 10^{-4}$ and
${\cal B}(\Xi_{bc}^+\to\Xi_{cc}^{++} D^-_s)\sim 10^{-3}$.
Therefore, we expect that the branching fractions of
${\bf B}_{bc} \to {\bf B}_c^{(\prime)}+ (M_c,M_{c\bar c})$ and
${\bf B}_{bc} \to {\bf B}_{cc} + (M,M_{\bar c})$ can be as large as
$(10^{-6},10^{-5},10^{-4},10^{-3})$, respectively.
According to the current luminosity at LHCb,
they are much more difficult to be measured experimentally.
The decayed meson particles are all easy to be identified.
In particular, by adopting the situation in Ref.~\cite{displaced},
where the weakly decaying double beauty hadrons have been discussed,
we take ${\bf B}_c^{(\prime)}$ and ${\bf B}_{cc}$ as the displaced baryons.
They travel sizeable distances before decaying,
which helps to distinguish the signal from the prompt background
that might come from the $c$-quark induced
${\bf B}_{bc}\to{\bf B}_b M({\bf B} M_b)$ decays~\cite{displaced}.
There may be some chances when more and more data are collected.
For example, with the subsequent $\Xi_{cc}^{++}$ decay,
the needed $\Xi_{bc}$ events to discovery $\Xi_{bc}^+\to\Xi_{cc}^{++}\pi^-$
are estimated to be around $10^9$.
We list the $SU(3)$ invariant decay amplitudes in Tables~\ref{tab3} and \ref{tab4} for completeness.
There are also several relations among different decay modes.

In conclusion, we have studied the two-body ${\bf B}_{bc}$ weak decays
using the $SU(3)$ flavor symmetry aiming to provide the most promising decay channels to discover ${\bf B}_{bc}$.
With the branching fractions estimated as a few times $10^{-2}$
for the Cabibbo-allowed $c$-quark weak decays,
the LHC may be able to discover ${\bf B}_{bc}$ given their excellent capabilities of identifying
${\bf B}^{(\prime)}$, $M_b$ and also reasonably good identifications for ${\bf B}_b$ baryons.
The decay modes ${\bf B}_{bc} \to {\bf B}_c^{(\prime)}+ (M_c,M_{c\bar c})$ and
${\bf B}_{bc} \to {\bf B}_{cc} + (M,M_{\bar c})$ induced by $b$-quark weak decay have
suppressed decay branching ratios as they are suppressed by CKM factor of $|V_{cb}/V_{cs}|\simeq 0.04$.
Nonetheless, their branching fractions are not necessarily too much smaller,
such as ${\cal B}(\Xi_{bc}^+\to\Xi_{cc}^{++} D^-_s)\simeq 10^{-3}$.
They may have some chance to be eventually detected at the LHCb.
We strongly urge our experimental colleagues to search for ${\bf B}_{bc}$ using two-body weak decays.

\section*{ACKNOWLEDGMENTS}
We would like to thank Prof.~Yue-Hong Xie, Prof.~Wenbin Qian and
Prof.~Eduardo Rodrigues for useful discussions.
XGH was supported in part by the MOST (Grant No. MOST 106-2112-M-002-003-MY3).
YKH was supported in part by NSFC (Grant No.~11675030).
This work was also supported in part by Key Laboratory for Particle Physics, Astrophysics and Cosmology, Ministry of Education, and Shanghai Key Laboratory for Particle Physics and Cosmology (Grant No. 15DZ2272100), and in part by the NSFC (Grant Nos. 11575111 and 11735010).

%
\begin{table}[h!]
\caption{Amplitudes of ${\bf B}_{bc}\to {\bf B}^{(\prime)} M_b$ with
$\lambda_{(ds,dd,sd)}=(V_{ud}V_{cs}^*,V_{ud}V_{cd}^*,V_{us}V_{cd}^*)$,
where $\lambda_{dd}=-\lambda_{ss}$ has been used.}\label{tab1}
\scriptsize
\begin{tabular}{|l|l|}
\hline
Decay modes& Amplitudes \\
\hline\hline
$\Xi_{bc}^0 \to \Xi^0 \bar B^0_s$
&$\lambda_{ds} d_2$ \\
$\Omega_{bc}^0 \to \Xi^0 \bar B^0$
&$\lambda_{ds} d_4$ \\
$\Xi_{bc}^0 \to \Sigma^+ B^-$
&$-\lambda_{ds}(d_1-d_2)$ \\
$\Xi_{bc}^+ \to \Sigma^+ \bar B^0$
&$-\lambda_{ds}(d_3-d_4)$ \\
$\Xi_{bc}^0 \to \Lambda \bar B^0$
&$\lambda_{ds}\frac{1}{\sqrt 6}(d_1+d_2+d_3+d_4)$ \\
$\Xi_{bc}^0 \to \Sigma^0 \bar B^0$
&$\lambda_{ds}\frac{1}{\sqrt 2}(d_1-d_2+d_3-d_4)$ \\
\hline\hline
$\Xi_{bc}^0 \to p B^-$
&$\lambda_{dd}(d_1-d_2)$ \\
$\Omega_{bc}^0 \to \Sigma^+ B^-$
&$\lambda_{dd}(d_1-d_2)$ \\
$\Omega_{bc}^0 \to \Xi^0 \bar B^0_s$
&$-\lambda_{dd} (d_2+d_4)$ \\
$\Xi_{bc}^0 \to n \bar B^0$
&$-\lambda_{dd}(d_2+d_4)$ \\
$\Xi_{bc}^+ \to p \bar B^0$
&$\lambda_{dd}(d_3-d_4)$ \\
$\Xi_{bc}^+ \to \Sigma^+ \bar B^0_s$
&$\lambda_{dd}(d_3-d_4)$ \\
$\Omega_{bc}^0 \to \Sigma^0 \bar B^0$
&$\lambda_{dd} \frac{-1}{\sqrt 2}(d_1-d_2+d_3)$ \\
$\Xi_{bc}^0 \to \Sigma^0 \bar B^0_s$
&$\lambda_{dd}\frac{-1}{\sqrt 2}(d_1+d_3-d_4)$ \\
$\Xi_{bc}^0 \to \Lambda \bar B^0_s$
&$\lambda_{dd}\frac{-1}{\sqrt 6}(d_1-2d_2+d_3+d_4)$ \\
$\Omega_{bc}^0 \to \Lambda \bar B^0$
&$\lambda_{dd}\frac{-1}{\sqrt 6}(d_1+d_2+d_3-2d_4)$ \\
\hline\hline
$\Omega_{bc}^0 \to n \bar B^0$
&$-\lambda_{sd} d_2$ \\
$\Xi_{bc}^0 \to n \bar B^0_s$
&$-\lambda_{sd} d_4$ \\
$\Omega_{bc}^0 \to p B^-$
&$\lambda_{sd}(d_1-d_2)$ \\
$\Omega_{bc}^0 \to \Sigma^0 \bar B^0_s$
&$\lambda_{sd}\frac{-1}{\sqrt 2}(d_1+d_3)$ \\
$\Xi_{bc}^+ \to p \bar B^0_s$
&$\lambda_{sd}(d_3-d_4)$ \\
$\Omega_{bc}^0 \to \Lambda \bar B^0_s$
&$\lambda_{sd}\frac{-1}{\sqrt 6}(d_1-2d_2+d_3-2d_4)$ \\

\hline
\end{tabular}
\begin{tabular}{|l|l|}
\hline
Decay modes& Amplitudes \\
\hline\hline
$\Xi_{bc}^0 \to \Sigma^{\prime+} B^-$
&$\lambda_{ds}\frac{1}{\sqrt 3}d'_1$ \\
$\Xi_{bc}^0 \to \Xi^{\prime0} \bar B^0_s$
&$\lambda_{ds}\frac{1}{\sqrt 3}d'_1$ \\
$\Omega_{bc}^0 \to \Xi^{\prime0} \bar B^0$
&$\lambda_{ds}\frac{1}{\sqrt 3}d'_2$ \\
$\Xi_{bc}^+ \to \Sigma^{\prime+} \bar B^0$
&$\lambda_{ds}\frac{1}{\sqrt 3}d'_2$ \\
$\Xi_{bc}^0 \to \Sigma^{\prime0} \bar B^0$
&$\lambda_{ds}\frac{1}{\sqrt 6}(d'_1+d'_2)$ \\[2mm]
\hline\hline
$\Xi_{bc}^0 \to \Delta^+ B^-$
&$\lambda_{dd}\frac{1}{\sqrt 3}d'_1$ \\
$\Omega_{bc}^0 \to \Sigma^{\prime+} B^-$
&$\lambda_{dd}\frac{-1}{\sqrt 3}d'_1$ \\
$\Xi_{bc}^+ \to \Delta^+ \bar B^0$
&$\lambda_{dd}\frac{1}{\sqrt 3}d'_2$ \\
$\Xi_{bc}^+ \to \Sigma^{\prime+} \bar B^0_s$
&$\lambda_{dd}\frac{-1}{\sqrt 3}d'_2$ \\
$\Xi_{bc}^0 \to \Delta^0 \bar B^0$
&$\lambda_{dd}\frac{1}{\sqrt 3}(d'_1+d'_2)$ \\
$\Omega_{bc}^0 \to \Xi^{\prime0} \bar B^0_s$
&$\lambda_{dd}\frac{-1}{\sqrt 3} (d'_1+d'_2)$ \\
$\Xi_{bc}^0 \to \Sigma^{\prime0}\bar B^0_s$
&$\lambda_{dd}\frac{1}{\sqrt 6}(d'_1-d'_2)$ \\
$\Omega_{bc}^0 \to \Sigma^{\prime0} \bar B^0$
&$\lambda_{dd}\frac{-1}{\sqrt 6}(d'_1-d'_2)$ \\
&\\[1mm]
\hline\hline
$\Omega_{bc}^0 \to \Delta^+ B^-$
&$\lambda_{sd}\frac{1}{\sqrt 3}d'_1$ \\
$\Omega_{bc}^0 \to \Delta^0 \bar B^0$
&$\lambda_{sd} \frac{1}{\sqrt 3}d'_1$ \\
$\Xi_{bc}^+ \to \Delta^+ \bar B^0_s$
&$\lambda_{sd}\frac{1}{\sqrt 3}d'_2$ \\
$\Xi_{bc}^0 \to \Delta^0 \bar B^0_s$
&$\lambda_{sd} \frac{1}{\sqrt 3} d'_2$ \\
$\Omega_{bc}^0 \to \Sigma^{\prime0} \bar B^0_s$
&$\lambda_{sd}\frac{1}{\sqrt 6}(d'_1+d'_2)$ \\[2mm]
\hline
\end{tabular}
\end{table}
%

%
\begin{table}[h!]
\caption{Amplitudes of ${\bf B}_{bc}\to {\bf B}_b^{(\prime)} M$.}\label{tab2}
\scriptsize
\begin{tabular}{|l|l|}
\hline
Decay modes& Amplitudes \\
\hline\hline
$\Xi_{bc}^0 \to \Xi_b^-\pi^+$
&$-\lambda_{ds}e_1$ \\
$\Xi_{bc}^+ \to \Xi_b^0 \pi^+$
&$-\lambda_{ds}e_3$ \\
$\Xi_{bc}^0 \to \Xi_b^0\pi^0$
&$\lambda_{ds}\frac{-1}{\sqrt 2}(e_1-e_3)$ \\
$\Omega_{bc}^0 \to \Xi_b^0 \bar K^0$
&$-\lambda_{ds} (e_3+e_4)$ \\
$\Xi_{bc}^0 \to \Xi_b^0\eta$
&$\lambda_{ds}\frac{-1}{\sqrt 6}(e_1+2e_2+e_3)$ \\
$\Xi_{bc}^0 \to \Lambda_b\bar K^0$
&$\lambda_{ds}(e_2-e_4)$ \\
&\\&\\&\\
&\\[2.5mm]
\hline\hline
$\Xi_{bc}^0 \to \Xi_b^- K^+$
&$\lambda_{dd} e_1$ \\
$\Omega_{bc}^0 \to \Xi_b^- \pi^+$
&$\lambda_{dd} e_1$ \\
$\Xi_{bc}^+ \to \Xi_b^0 K^+$
&$\lambda_{dd}e_3$ \\
$\Xi_{bc}^+ \to \Lambda_b \pi^+$
&$-\lambda_{dd}e_3$ \\
$\Xi_{bc}^0 \to \Xi_b^0 K^0$
&$\lambda_{dd}(e_2+e_3)$ \\
$\Omega_{bc}^0 \to \Lambda_b^0 \bar K^0$
&$-\lambda_{dd} (e_2+e_3)$ \\
$\Omega_{bc}^0 \to \Xi_b^0 \pi^0$
&$\lambda_{dd} \frac{1}{\sqrt 2}(e_1+e_4)$ \\
$\Xi_{bc}^0 \to \Lambda_b\pi^0$
&$\lambda_{dd}\frac{-1}{\sqrt 2}(e_1+e_2-e_3-e_4)$ \\
$\Xi_{bc}^0 \to \Lambda_b\eta$
&$\lambda_{dd}\frac{-1}{\sqrt 6}(e_1-e_2+e_3+3e_4)$ \\
$\Omega_{bc}^0 \to \Xi_b^0 \eta$
&$\lambda_{dd}\frac{1}{\sqrt 6}(e_1+2 e_2-2 e_3-3 e_4)$\\
&\\
&\\
&\\
&\\
&\\
&\\
&\\
&\\[1mm]
\hline\hline
$\Omega_{bc}^0 \to \Xi_b^- K^+$
&$\lambda_{sd}e_1$ \\
$\Xi_{bc}^+ \to \Lambda_b K^+$
&$-\lambda_{sd}e_3$ \\
$\Omega_{bc}^0 \to \Lambda_b \pi^0$
&$\lambda_{sd}\frac{-1}{\sqrt 2}(e_1+e_2)$ \\
$\Omega_{bc}^0 \to \Xi_b^0 K^0$
&$\lambda_{sd}(e_2- e_4)$ \\
$\Xi_{bc}^0 \to \Lambda_b K^0$
&$-\lambda_{sd} (e_3+e_4)$ \\
$\Omega_{bc}^0 \to \Lambda_b \eta$
&$\lambda_{sd}\frac{-1}{\sqrt 6}(e_1-e_2-2 e_3)$ \\
&\\
&\\
&\\
&\\[3mm]
\hline
\end{tabular}
\begin{tabular}{|l|l|}
\hline
Decay modes& Amplitudes \\
\hline\hline
$\Xi_{bc}^0 \to \Omega_b^{-} K^+$
&$\lambda_{ds} e'_1$ \\
$\Xi_{bc}^0 \to \Sigma_b^{+}K^-$
&$\lambda_{ds}e'_2$ \\
$\Xi_{bc}^0 \to \Xi_b^{\prime0} \eta$
&$\lambda_{ds} \frac{1}{2\sqrt 3}(e'_1-2e'_2+e'_3)$ \\
$\Xi_{bc}^+ \to \Sigma_b^{+}\bar K^0$
&$\lambda_{ds}e'_4$ \\
$\Omega_{bc}^0 \to \Omega_b^- \pi^+$
&$\lambda_{ds} e'_5$ \\
$\Xi_{bc}^0 \to \Xi_b^{\prime0} \pi^0$
&$\lambda_{ds} \frac{1}{2}(e'_1-e'_3)$ \\
$\Xi_{bc}^+ \to \Xi_b^{\prime0} \pi^+$
&$\lambda_{ds} \frac{1}{\sqrt 2}(e'_3+e'_5)$ \\
$\Xi_{bc}^0 \to \Xi_b^{\prime-} \pi^+$
&$\lambda_{ds} \frac{1}{\sqrt 2}(e'_1+e'_5)$ \\
$\Xi_{bc}^0 \to \Sigma_b^{0}\bar K^0$
&$\lambda_{ds}\frac{1}{\sqrt 2} (e'_2+e'_4)$ \\
$\Omega_{bc}^0 \to \Xi_b^{\prime0}\bar K^0$
&$\lambda_{ds}\frac{1}{\sqrt 2} (e'_3+e'_4)$ \\
\hline\hline
$\Xi_{bc}^0 \to \Sigma_b^{+} \pi^-$
&$\lambda_{dd} e'_2$ \\
$\Omega_{bc}^0 \to \Sigma_b^{+} K^-$
&$-\lambda_{dd}e'_2$\\
$\Xi_{bc}^+ \to \Sigma_b^{+} \pi^0$
&$\lambda_{dd} \frac{-1}{\sqrt 2}e'_4$ \\
$\Xi_{bc}^+ \to \Sigma_b^{+} \eta$
&$\lambda_{dd} \sqrt\frac{3}{ 2}e'_4$ \\
$\Omega_{bc}^0 \to \Xi_b^{\prime0} \pi^0$
&$-\lambda_{dd}\frac{1}{2} (e'_1+ e'_4)$ \\
$\Xi_{bc}^0 \to \Sigma_b^{-} \pi^+$
&$\lambda_{dd} (e'_1+e'_5)$ \\
$\Omega_{bc}^0 \to \Omega_b^{-} K^+$
&$-\lambda_{dd} (e'_1+ e'_5)$ \\
$\Xi_{bc}^0 \to \Xi_b^{\prime-} K^+$
&$\lambda_{dd}\frac{1}{\sqrt 2}(e'_1-e'_5)$ \\
$\Omega_{bc}^0 \to \Xi_b^{\prime-} \pi^+$
&$\lambda_{dd} \frac{-1}{\sqrt 2}(e'_1- e'_5)$ \\
$\Omega_{bc}^0 \to \Sigma_b^{0} \bar K^0$
&$\lambda_{dd}\frac{-1}{\sqrt 2}( e'_2-e'_3)$ \\
$\Xi_{bc}^0 \to \Xi_b^{\prime0}K^0$
&$\lambda_{dd}\frac{1}{\sqrt 2}(e'_2-e'_3)$ \\
$\Xi_{bc}^+ \to \Sigma_b^{0} \pi^+$
&$\lambda_{dd} \frac{1}{\sqrt 2}(e'_3+e'_5)$ \\
$\Xi_{bc}^+ \to \Xi_b^{\prime0} K^+$
&$\lambda_{dd} \frac{-1}{\sqrt 2}(e'_3+e'_5)$ \\
$\Xi_{bc}^0 \to \Sigma_b^{0}\pi^0$
&$\lambda_{dd}\frac{1}{2}(e'_1-e'_2-e'_3-e'_4)$ \\
$\Xi_{bc}^0 \to \Sigma_b^{0}\eta$
&$\lambda_{dd}\frac{1}{2\sqrt 3}(e'_1+e'_2+e'_3+3e'_4)$ \\
$\Omega_{bc}^0 \to \Xi_b^{\prime0} \eta$
&$\lambda_{dd} \frac{-1}{2\sqrt 3}(e'_1-2e'_2-2e'_3-3e'_4)$ \\
\hline\hline
$\Omega_{bc}^0 \to \Sigma_b^{-} \pi^+$
&$\lambda_{sd}e'_1$ \\
$\Omega_{bc}^0 \to \Sigma_b^{+} \pi^-$
&$\lambda_{sd}e'_2$ \\
$\Xi_{bc}^+ \to \Sigma_b^{+} K^0$
&$\lambda_{sd} e'_4$ \\
$\Xi_{bc}^0 \to \Sigma_b^{-} K^+$
&$\lambda_{sd} e'_5$ \\
$\Omega_{bc}^0 \to \Sigma_b^{0} \pi^0$
&$\lambda_{sd} \frac{1}{2}(e'_1-e'_2)$ \\
$\Omega_{bc}^0 \to \Xi_b^{\prime-} K^+$
&$\lambda_{sd} \frac{1}{\sqrt 2}(e'_1+e'_5)$ \\
$\Omega_{bc}^0 \to \Xi_b^{\prime0} K^0$
&$\lambda_{sd} \frac{1}{\sqrt 2}(e'_2+e'_4)$ \\
$\Xi_{bc}^0 \to \Sigma_b^{0} K^0$
&$\lambda_{sd} \frac{1}{\sqrt 2}(e'_3+e'_4)$ \\
$\Xi_{bc}^+ \to \Sigma_b^{0} K^+$
&$\lambda_{sd} \frac{1}{\sqrt 2}(e'_3+e'_5)$ \\
$\Omega_{bc}^0 \to \Sigma_b^{0} \eta$
&$\lambda_{sd} \frac{1}{2\sqrt 3}(e'_1+e'_2-2e'_3)$ \\
\hline
\end{tabular}
\end{table}

\newpage
%
\begin{table}[h!]
\caption{Amplitudes of ${\bf B}_{bc}\to {\bf B}_{cc} M$ and
${\bf B}_{bc}\to {\bf B}_c^{(\prime)} M_c$ with
$\lambda_{ud(s)}^c=V_{cb}V_{ud(s)}^*$.}\label{tab3}
\scriptsize
\begin{tabular}{|l|l|}
\hline
Decay modes& Amplitudes \\
\hline\hline
$\Xi_{bc}^+ \to \Omega_{cc}^+ K^0$
&$\lambda_{ud}^ca_1$ \\
$\Xi_{bc}^+ \to \Xi_{cc}^{+} \bar K^0$
&$\lambda_{us}^{c}a_1$ \\
$\Omega_{bc}^0 \to \Omega_{cc}^{+} \pi^-$
&$\lambda_{ud}^c a_2$ \\
$\Xi_{bc}^0 \to \Xi_{cc}^{+} K^-$
&$\lambda_{us}^{c}a_2$ \\
$\Omega_{bc}^0 \to \Xi_{cc}^{+} K^-$
&$\lambda_{ud}^c a_3$ \\
$\Xi_{bc}^0 \to \Omega_{cc}^+ \pi^-$
&$\lambda_{us}^{c}a_3$ \\
$\Xi_{bc}^+ \to \Omega_{cc}^+ \pi^0$
&$\lambda_{us}^{c}{1\over \sqrt{2}}a_3$ \\
$\Xi_{bc}^+ \to \Xi_{cc}^{++} \pi^-$
&$\lambda_{ud}^c(a_1+a_2)$ \\
$\Xi_{bc}^+ \to \Xi_{cc}^{++} K^-$
&$\lambda_{us}^{c}(a_1+a_2)$ \\
$\Xi_{bc}^0 \to \Xi_{cc}^{+} \pi^-$
&$\lambda_{ud}^c(a_2+a_3)$ \\
$\Omega_{bc}^0 \to \Omega_{cc}^+ K^-$
&$\lambda_{us}^{c}(a_2+a_3)$\\
$\Xi_{bc}^+ \to \Xi_{cc}^{+} \eta$
&$\lambda_{ud}^c{1\over \sqrt{6}}(a_1+a_3)$ \\
$\Xi_{bc}^+ \to \Xi_{cc}^{+} \pi^0$
&$\lambda_{ud}^c{-1\over \sqrt{2}}(a_1-a_3)$ \\
$\Xi_{bc}^+ \to \Omega_{cc}^{+} \eta$
&$\lambda_{us}^{c}{-1\over \sqrt{6}}(2a_1-a_3)$ \\
&\\
& \\[3mm]
\hline
\end{tabular}
\begin{tabular}{|l|l|}
\hline
Decay modes& Amplitudes \\
\hline\hline
$\Xi_{bc}^+ \to \Xi_{c}^{0} D_s^+$
&$\lambda_{ud}^{c}b_1$ \\
$\Xi_{bc}^+ \to \Xi_{c}^{0} D^+$
&$-\lambda_{us}^{c}b_1$ \\
$\Omega_{bc}^0 \to \Xi_{c}^{0} D^0$
&$\lambda_{ud}^{c}b_2$ \\
$\Xi_{bc}^0 \to \Xi_{c}^{0} D^0$
&$-\lambda_{us}^{c}b_2$ \\
$\Xi_{bc}^+ \to \Lambda_{c}^{+} D^0$
&$-\lambda_{ud}^{c}(b_1+b_2)$ \\
$\Xi_{bc}^+ \to \Xi_{c}^{+} D^0$
&$-\lambda_{us}^{c}(b_1+b_2)$ \\
\hline\hline
$\Xi_{bc}^+ \to \Sigma_{c}^{0} D^+$
&$\lambda_{ud}^{c}b'_1$ \\
$\Xi_{bc}^+ \to \Omega_{c}^{0} D_s^+$
&$\lambda_{us}^{c}b'_1$ \\
$\Xi_{bc}^+ \to \Xi_{c}^{\prime0} D_s^+$
&$\lambda_{ud}^{c}\frac{1}{\sqrt 2}b'_1$ \\
$\Xi_{bc}^+ \to \Xi_{c}^{\prime0} D^+$
&$\lambda_{us}^{c}\frac{1}{\sqrt 2}b'_1$ \\
$\Xi_{bc}^0 \to \Sigma_{c}^{0} D^0$
&$\lambda_{ud}^{c}b'_2$ \\
$\Omega_{bc}^0 \to \Xi_{c}^{\prime0} D^0$
&$\lambda_{ud}^{c}\frac{1}{\sqrt 2}b'_2$ \\
$\Omega_{bc}^0 \to \Omega_{c}^{0} D^0$
&$\lambda_{us}^{c}b'_2$ \\
$\Xi_{bc}^0 \to \Xi_{c}^{\prime0} D^0$
&$\lambda_{us}^{c} \frac{1}{\sqrt 2}b'_2$ \\
$\Xi_{bc}^+ \to \Sigma_{c}^{+} D^0$
&$\lambda_{ud}^{c}\frac{1}{\sqrt 2}(b_1^{\prime}+b_2^{\prime})$ \\
$\Xi_{bc}^+ \to \Xi_{c}^{\prime+} D^0$
&$\lambda_{us}^{c}\frac{1}{\sqrt 2}(b'_1+b'_2)$ \\
\hline
\end{tabular}
\end{table}
%

%
\begin{table}[h!]
\caption{Amplitudes of ${\bf B}_{bc}\to {\bf B}_{cc}  M_{\bar c}$ and
${\bf B}_{bc}\to {\bf B}_c^{(\prime)} M_{c\bar c}$ with
$\lambda_{cd(s)}^c=V_{cb}V_{cd(s)}^*$.}\label{tab4}
\scriptsize
\begin{tabular}{|l|l|}
\hline
Decay modes& Amplitudes \\
\hline\hline
$\Xi_{bc}^+ \to \Xi_{cc}^{++} D^-$
&$\lambda_{cd}^{c} f_1$ \\
$\Omega_{bc}^0 \to \Omega_{cc}^{+} D^-$
&$\lambda_{cd}^{c} f_1$ \\

$\Xi_{bc}^+ \to \Xi_{cc}^{++} D^-_s$
&$\lambda_{cs}^{c} f_1$ \\
$\Xi_{bc}^0 \to \Xi_{cc}^{+} D^-_s$
&$\lambda_{cs}^{c} f_1$ \\
$\Xi_{bc}^+ \to \Xi_{c}^{+} \bar D^0$
&$\lambda_{cd}^{c} f_2$ \\
$\Omega_{bc}^0 \to \Xi_{cc}^{+} D^-_s$
&$\lambda_{cd}^{c} f_2$ \\

$\Xi_{bc}^+ \to \Omega_{cc}^+ \bar D^0$
&$\lambda_{cs}^{c} f_2$ \\
$\Xi_{bc}^0 \to \Omega_{cc}^+ D^-$
&$\lambda_{cs}^{c} f_2$ \\
$\Xi_{bc}^0 \to \Xi_{cc}^{+} D^-$
&$\lambda_{cd}^{c} (f_1+f_2)$ \\
$\Omega_{bc}^0 \to \Omega_{cc}^+ D^-_s$
&$\lambda_{cs}^{c}(f_1+f_2)$ \\[2mm]
\hline
\end{tabular}
\begin{tabular}{|l|l|}
\hline
Decay modes& Amplitudes \\
\hline\hline
$\Xi_{bc}^+ \to \Lambda_{c}^{+} M_{c\bar c}$
&$\lambda_{cd}^{c} g_{c\bar c}$ \\
$\Omega_{bc}^0 \to \Xi_{c}^{0} M_{c\bar c}$
&$-\lambda_{cd}^{c} g_{c\bar c}$ \\
$\Xi_{bc}^+ \to \Xi_{c}^{+} M_{c\bar c}$
&$\lambda_{cs}^{c} g_{c\bar c}$ \\
$\Xi_{bc}^0 \to \Xi_{c}^{0} M_{c\bar c}$
&$\lambda_{cs}^{c} g_{c\bar c}$ \\
\hline
$\Xi_{bc}^+ \to \Sigma_{c}^{+} M_{c\bar c}$
&$\lambda_{cd}^{c} \frac{1}{\sqrt 2}g'_{c\bar c}$ \\
$\Xi_{bc}^0 \to \Sigma_{c}^{0} M_{c\bar c}$
&$\lambda_{cd}^{c}g'_{c\bar c}$ \\
$\Omega_{bc}^0 \to \Xi_{c}^{\prime0} M_{c\bar c}$
&$\lambda_{cd}^{c}\frac{1}{\sqrt 2}g'_{c\bar c}$ \\
$\Xi_{bc}^+ \to \Xi_{c}^{\prime+} M_{c\bar c}$
&$\lambda_{cs}^{c}\frac{1}{\sqrt 2}g'_{c\bar c}$ \\
$\Xi_{bc}^0 \to \Xi_{c}^{\prime0} M_{c\bar c}$
&$\lambda_{cs}^{c} \frac{1}{\sqrt 2}g'_{c\bar c}$ \\
$\Omega_{bc}^0 \to \Omega_{c}^{0} M_{c\bar c}$
&$\lambda_{cs}^{c} g'_{c\bar c}$ \\
\hline
\end{tabular}
\end{table}
%
\begin{table}[h!]
\caption{Needed $\Xi_{bc}$ events for the Cabibbo-allowed
$\Xi_{bc}\to {\bf B}M_b$ and $\Xi_{bc}\to{\bf B}_b M$ decays.}\label{tab5}
\scriptsize
\begin{tabular}{|l|l|c|}
\hline
Decay modes&$\;\;\;\;\;\;$Subsequent decays~\cite{pdg,Hsiao:2015cda}&needed $\Xi_{bc}$ events\\
\hline\hline
$\Xi_{bc}^0 \to \Xi^0 \bar B^0_s$
&${\cal B}(\Xi^0\to \pi^0(\Lambda\to) p \pi^-)\simeq 64\%$,
&$10^7$\\
&${\cal B}(\bar B^0_s\to \rho^-(D^+_s\to)\pi^+\pi^+\pi^-)\simeq 7.5 \times 10^{-5}$
&\\\hline
$\Xi_{bc}^0 \to \Sigma^+ B^-$
&${\cal B}(\Sigma^+\to p\pi^0)\simeq 52\%$,
&$10^{7}$\\
&${\cal B}(B^-\to \rho^-(D^0\to)\pi^+\pi^-)\simeq 2\times 10^{-5}$
&\\\hline
$\Xi_{bc}^+ \to \Sigma^+ \bar B^0$
&${\cal B}(\Sigma^+\to p\pi^0)\simeq 52\%$,
&$10^{8}$\\
&${\cal B}(\bar B^0\to \pi^-(D^+\to)\pi^+\pi^0)\simeq 3\times 10^{-6}$
&\\\hline
$\Xi_{bc}^0 \to \Lambda \bar B^0$
&${\cal B}(\Lambda\to p\pi^-)\simeq 64\%$,
&$10^{8}$\\
&${\cal B}(\bar B^0\to \pi^-(D^+\to)\pi^+\pi^0)\simeq 3\times 10^{-6}$
&\\\hline
$\Xi_{bc}^0 \to \Sigma^0 \bar B^0$
&${\cal B}(\Sigma^0\to \gamma(\Lambda\to)p\pi^-)\simeq 64\%$,
&$10^{8}$\\
&${\cal B}(\bar B^0\to \pi^-(D^+\to)\pi^+\pi^0)\simeq 3\times 10^{-6}$
&\\
\hline\hline
$\Xi_{bc}^0 \to \Xi_b^-\pi^+$
&${\cal B}(\Xi_b^-\to \Xi^- J/\Psi)=10^{-4}-10^{-3}$,
&$10^{7-8}$\\
&${\cal B}(\Xi^-\to \pi^-(\Lambda\to) p \pi^-)\simeq 64\%$,
&\\
&${\cal B}(J/\Psi\to \ell^+\ell^-)\simeq 12\%$
&\\\hline
$\Xi_{bc}^+ \to \Xi_b^0 \pi^+$
&${\cal B}(\Xi_b^0\to \Xi^0 J/\Psi)=10^{-4}-10^{-3}$,
&$10^{7-8}$\\
&${\cal B}(\Xi^0\to \pi^0(\Lambda\to) p \pi^-)\simeq 64\%$,
&\\
&${\cal B}(J/\Psi\to \ell^+\ell^-)\simeq 12\%$
&\\\hline
$\Xi_{bc}^0 \to \Xi_b^0\pi^0$
&${\cal B}(\Xi_b^0\to \Xi^0 J/\Psi)=10^{-4}-10^{-3}$,
&$10^{7-8}$\\
&${\cal B}(\Xi^0\to \pi^0(\Lambda\to )p \pi^-)\simeq 64\%$,
&\\
&${\cal B}(J/\Psi\to \ell^+\ell^-)\simeq 12\%$
&\\\hline
$\Xi_{bc}^0 \to \Xi_b^0\eta$
&${\cal B}(\Xi_b^0\to \Xi^0 J/\Psi)=10^{-4}-10^{-3}$,
&$10^{7-8}$\\
&${\cal B}(\Xi^0\to \pi^0(\Lambda\to )p \pi^-)\simeq 64\%$,
&\\
&${\cal B}(J/\Psi\to \ell^+\ell^-)\simeq 12\%$,
&\\
&${\cal B}(\eta\to \pi^+\pi^-\pi^0)\simeq 23\%$
&\\\hline
$\Xi_{bc}^0 \to \Lambda_b\bar K^0$
&${\cal B}(\Lambda_b\to \Lambda J/\Psi)=10^{-4}-10^{-3}$,
&$10^{7-8}$\\
&${\cal B}(\Lambda\to p\pi^-)\simeq 64\%$,
&\\
&${\cal B}(J/\Psi\to \ell^+\ell^-)\simeq 12\%$,
&\\
&${\cal B}(\bar K^0_s\to \pi\pi)\simeq 100\%$
&\\\hline
\end{tabular}
\end{table}

\end{document}